\documentclass{iaus}
\usepackage{graphicx}

\title[short title of paper]{Galaxy-galaxy lensing studies from COMBO-17}

\author[Kleinheinrich et al.]{Martina Kleinheinrich$^1$, Hans-Walter
  Rix$^1$, Peter Schneider$^2$, Thomas Erben$^2$, Klaus Meisenheimer$^1$,
  Christian Wolf~$^3$, \and Mischa Schirmer$^4$} 
\affiliation{$^1$Max-Planck-Institut f\"ur Astronomie, K\"onigstuhl 17, D-69117 Heidelberg, Germany\\[\affilskip]
$^2$Institut f\"ur Astrophysik und Extraterrestrische Forschung, Universit\"at
  Bonn, Auf dem H\"ugel 71, D-53121 Bonn, Germany\\[\affilskip]
$^3$Department of Physics, Denys Wilkinson Bldg., University of Oxford, Keble
  Road, Oxford, OX1 3RH, U.K.\\[\affilskip] $^4$Isaac Newton Group of
  Telescopes, Apartado de correos 321, E-38700 Santa Cruz de La Palma, Spain}

\pubyear{2004}
\volume{225}
\pagerange{1--8}
\date{?? and in revised form ??}
\jname{Impact of Gravitational Lensing on Cosmology}
\editors{Mellier, Y. \& Meylan,G. eds.}
\begin{document}

\maketitle

\begin{abstract}
We study the dark matter halos of galaxies with galaxy-galaxy lensing using
the COMBO-17 survey. This survey offers an unprecedented data set for 
studying lens galaxies at $z=0.2-0.7$ including redshift information and
spectral classification from 17 optical filters for objects brighter than
$R=24$. So far, redshifts and classification for the lens galaxies have mainly
been available for local surveys like the Sloan Digital Sky Survey
(SDSS). Further, redshifts for the source galaxies have typically not been
available at all but had to be estimated from redshift probability
distribution which -- for faint surveys -- even had to be extrapolated.

To study the dark matter halos we parametrize the lens galaxies as singular 
isothermal spheres (SIS) or by Navarro-Frenk-White (NFW) profiles. In both
cases we find a dependence of the velocity dispersion or virial radius,
respectively, on lens luminosity and colour. For the SIS model, we are able to
reproduce the Tully-Fisher/Faber-Jackson relation on a scale of 
$150h^{-1}~\mathrm{kpc}$. For the NFW profile we also calculate virial masses,
mass-to-light ratios and rotation velocities.

Finally, we investigate differences between the three survey fields used here.
\end{abstract}

\firstsection 
\section{Outline of the method}
Galaxy-galaxy lensing uses the distortions of background galaxies to study
the mass distribution around foreground galaxies. In a typical lens situation,
the shear from a foreground lens is only weak. Therefore, galaxy-galaxy
lensing can only study dark matter halos of galaxies statistically by
averaging over thousands of lens galaxies. For reviews on galaxy-galaxy
lensing see \cite{mellier1999} and \cite{bartelmann2001}.

We use the maximum-likelihood technique proposed by \cite{schneider1997}.
First, we have to identify lenses and source galaxies which we do based on
accurate photometric redshifts. Next, we adopt a specific lens model to
calculate for each background galaxy the shear contributions from each
foreground galaxy within a certain annulus. The estimated shear is compared to
the observed shapes of the sources for a range of input parameters of the lens
model and those parameters which maximize the likelihood are determined. Here,
we use the singular isothermal sphere (SIS) and the Navarro-Frenk-White (NFW) 
profile to model the lenses.

We adopt $(\Omega_m,\Omega_\Lambda)=(0.3,0.7)$ and
$H_0=100h~\mathrm{km~s}^{-1}\mathrm{Mpc}^{-1}$.

\section{Data: The COMBO-17 survey}\label{sec:data}
For our investigation we use the COMBO-17 survey (\cite{wolf2004}) which
is a deep survey with very good imaging quality and accurate photometric
redshifts. All data are taken with the Wide Field Imager at the MPG/ESO 2.2-m
telescope on La Silla, Chile. The survey consists of 4 fields of which 3 are
used here. The limiting magnitude is $R\approx 25.5$. Deep $R$-band
observations were taken in the best seeing conditions (below 0.8"
PSF). Observations in $UBVRI$ and 12 medium-band filters are used to derive
restframe colours and accurate photometric redshifts with
$\sigma_z<0.1$ at $R<24$ and $\sigma_z<0.01$ at $R<21$. These allow us to
select both lenses and sources based on their redshifts and to select and
study subsamples of lens galaxies based on their restframe colours.

\section{Results}
In the following, sources are all galaxies with $R=18-24$ and
$z_\mathrm{s}=0.3-1.4$. Lenses are galaxies with $R=18-24$,
$z_\mathrm{d}=0.2-0.7$. The shear of a specific lens galaxy on a specific
source galaxy is only considered if $z_\mathrm{d}<z_\mathrm{s}-0.1$ for that
lens-source pair. Further, the projected separation between lens and source
must be smaller than $r_\mathrm{max}=150 h^{-1}~\mathrm{kpc}$ at the redshift
of the lens when the lens is modelled as SIS. At this $r_\mathrm{max}$ we
obtain the tightest constraints. When modelling lenses by NFW
profiles we extend the maximum separation to $r_\mathrm{max}=400
h^{-1}~\mathrm{kpc}$ to ensure that the region around the virial radius is
probed. Due to the size of galaxy images, a minimum angular separation of 8"
between lenses and sources is required in order to avoid that shape
measurements of sources are biased by the light of the lenses.

We investigate the lens sample as a whole and additionally split it into red
and blue subsamples based on restframe colours. Galaxies with $<U-V>\leq
1.15-0.31\times z-0.08(M_V-5~log~h +20)$ define the blue sample while all
other galaxies are in the red sample (\cite{bell2004}).

\subsection{SIS and Tully-Fisher/Faber-Jackson relation}
\label{sect:sis}
The density profile of the SIS is given by
$\rho(r)=\sigma_\mathrm{v}^2/2\pi G r^2$ where $\sigma_\mathrm{v}$ is
the velocity dispersion and $r$ is the distance from the center of the
lens. We assume that the velocity dispersion depends on the luminosity of a
galaxy, $\sigma_\mathrm{v}/\sigma_\star=(L/L_\star)^\eta$, where
$L_\star=10^{10}h^{-2}L_\odot$ is a characteristic luminosity measured in the
restframe SDSS $r$-band. This is the Tully-Fisher or Faber-Jackson relation
which was derived for galaxies on smaller scales than used here from their
rotation curves or stellar velocity dispersion.
 
\begin{figure}[ht!]
  \begin{center}
  \leavevmode
  \begin{minipage}[l]{0.49\textwidth}
    \includegraphics[width=\hsize,angle=-90]{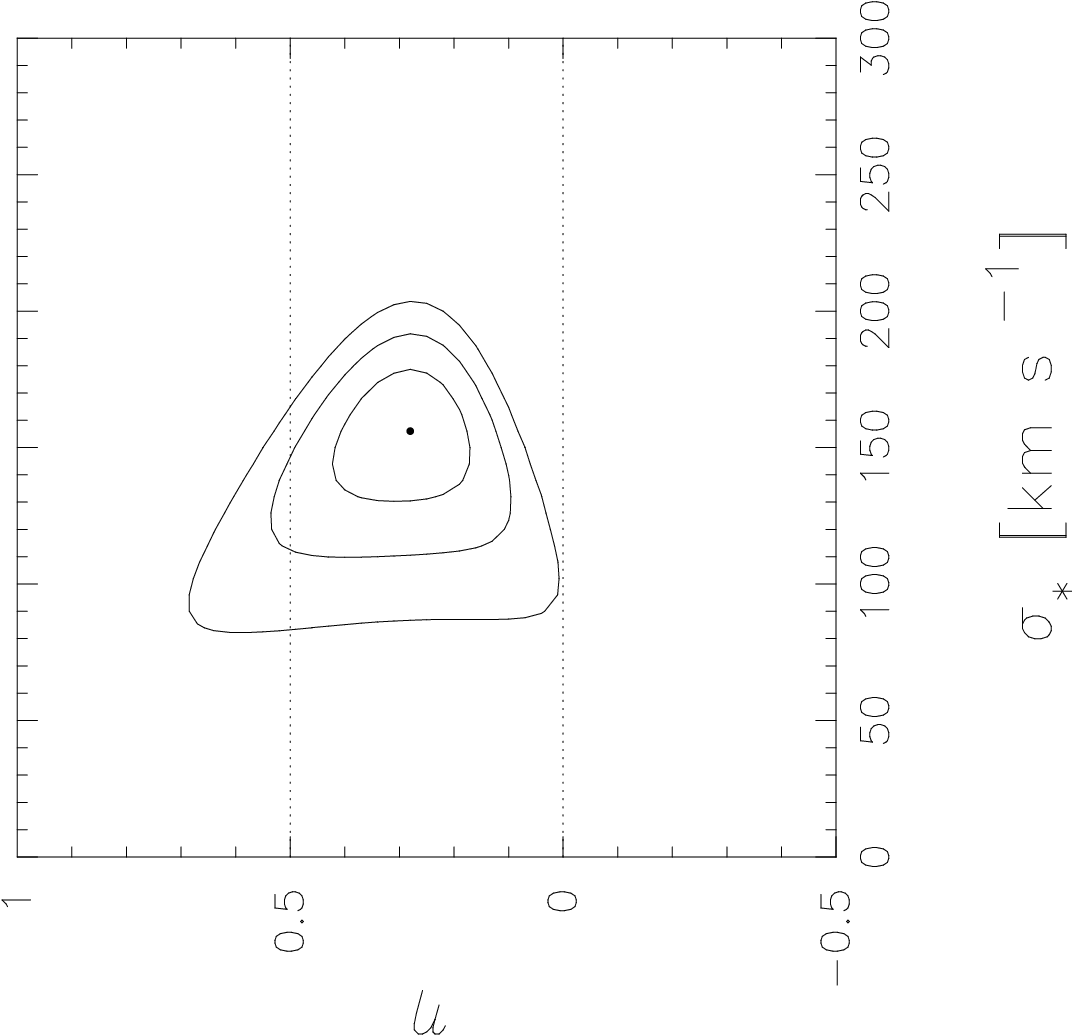}
  \end{minipage}
  \begin{minipage}[r]{0.49\textwidth}
    \includegraphics[width=\hsize,angle=-90]{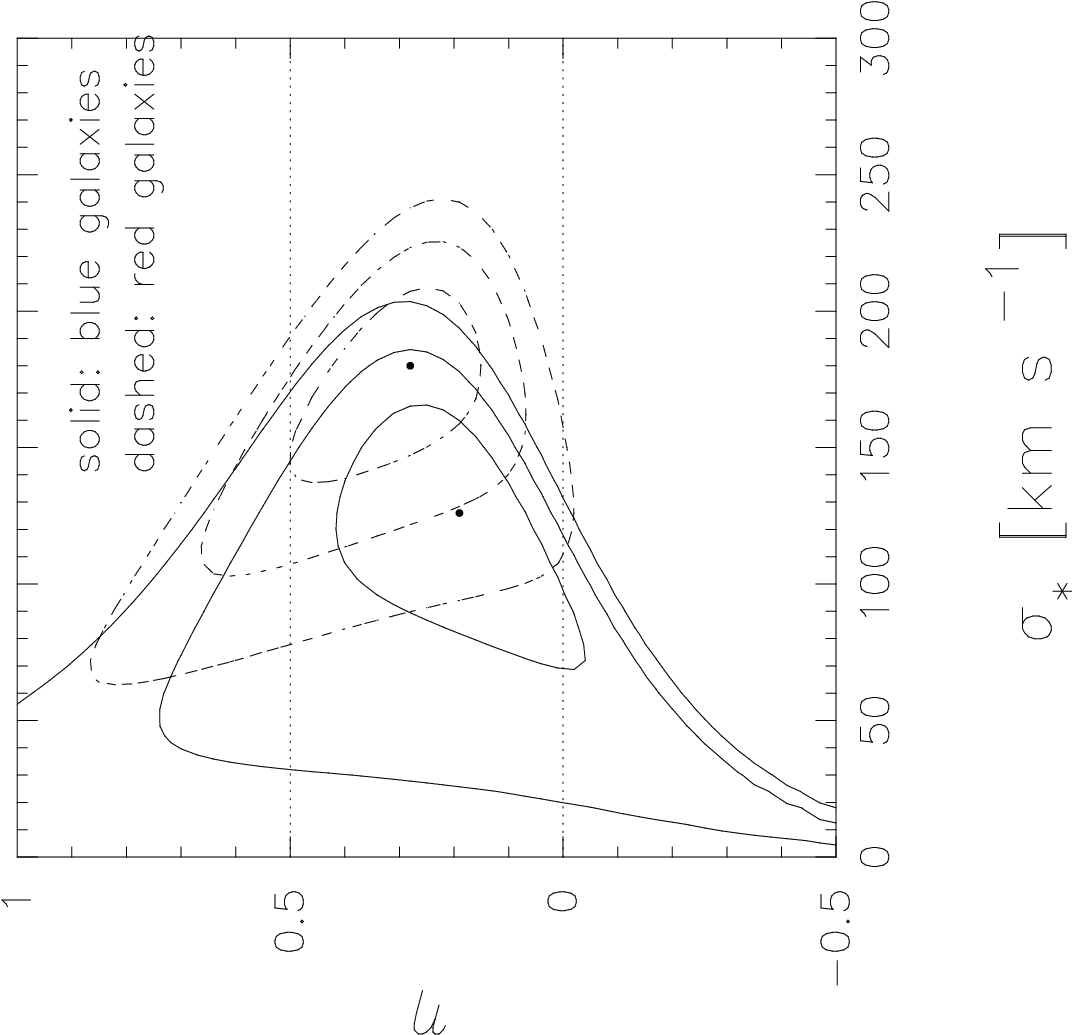}
  \end{minipage}
  \end{center}
  \caption{Constraints on dark matter halos modelled as SIS. The left panel
    uses the whole lens sample. In the right panel, the lens sample is split
    into blue and red lenses.}
  \label{sis}
\end{figure}
The left panel of Figure \ref{sis} shows 1-, 2- and 3-$\sigma$ contours for
the model parameters $\sigma_\star$ and $\eta$ derived for the whole lens
sample. The best-fit parameters with 1-$\sigma$ errors are
$\sigma_\star=156^{+18}_{-18}~\mathrm{km/s}$ and
$\eta=0.28^{+0.12}_{-0.09}$. These values agree very well with expectations
from e.g.\ rotation curve measurements.

We want to compare this result to the galaxy-galaxy lensing measurement from
the Red-Sequence Cluster Survey (RCS, \cite{hoekstra2004}) which probes lens
galaxies in a comparable redshift range and uses a similar
modelling. \cite{hoekstra2004} find $\sigma_\star=140\pm 4~\mathrm{km/s}$ at 
fixed $\eta=0.3$. However, \cite{hoekstra2004} use a characteristic luminosity
of $L_B=10^{10}h^{-2}L_\odot$ measured in the $B$-band instead of the $r$-band
as we do here. From the restframe luminosities of galaxies in COMBO-17 we
estimate that galaxies with $L_B=10^{10}h^{-2}L_\odot$ have $L_r=1.25\times
10^{10}h^{-2}L_\odot$. Further, \cite{hoekstra2004} use pairs with projected
separations up to 2' corresponding to about
$r_\mathrm{max}=350h^{-1}~\mathrm{kpc}$. Although for the SIS model the
velocity dispersion should be independent of radius, we find a decline of
the fitted $\sigma_\star$ with increasing $r_\mathrm{max}$. Using
$L_\star=1.25\times 10^{10}h^{-2}L_\odot$ and
$r_\mathrm{max}=350h^{-1}~\mathrm{kpc}$ we measure
$\sigma_\star=138^{+18}_{-24}~\mathrm{km/s}$ and $\eta=0.34^{+0.18}_{-0.12}$
in very good agreement with the RCS result. 

The error on $\sigma_\star$ is about 5 times smaller for the RCS than for
COMBO-17. Given the about 60 times larger area of the RCS this 
is not surprising. The uncertainties should also be influenced by the qualitiy
of the redshift information. The measurement from the RCS uses observations in
a single passband only and does therefore not have redshift estimates for
individual objects. In \cite{kleinheinrich2004} we find that the velocity
dispersion can be well constrained even in the absence of redshift
information. Redshifts for individual lens galaxies reduce the errors on
$\sigma_\star$ by only 15\%. However, they are essential for measuring the
dependence of the velocity dispersion (or mass) on luminosity -- the errors on
$\eta$ increase by a factor of 2.5 when omitting the lens
redshifts. Individual redshifts for the sources are not important as long as
the redshift distribution is known.

The right panel of Figure \ref{sis} shows likelihood contours for the blue and
red subsamples. While no significant change in $\eta$ is seen, a 2-$\sigma$
difference in $\sigma_\star$ is seen between the two lens populations. The
best-fit velocity dispersions are $\sigma_\star=126^{+30}_{-36}~\mathrm{km/s}$
for the blue sample and $\sigma_\star=180^{+24}_{-30}~\mathrm{km/s}$ for the
red sample, respectively. The red sample consists of 2579 galaxies, the blue
sample of 9898 galaxies. Although only about 20\% of the lenses are red, this
subsample gives even tighter constraints than the blue subsample. This shows
clearly that most of the galaxy-galaxy lensing signal comes from red galaxies.

\subsection{NFW and 'Tully-Fisher/Faber-Jackson' relation}
Next, we model lens galaxies by NFW profiles. The density profile is given by
$\rho(r)=\delta_c/(r/r_s(1+r/r_s)^2)$. $r_s$ is a characteristic
scale radius at which the density profile changes from $\rho(r)\propto r^{-1}$
to $\rho(r)\propto r^{-3}$. $\delta_c$ is related to the concentration
$c$. The virial radius $r_\mathrm{vir}$ is defined by
$r_\mathrm{vir}=r_sc$. Here, the virial radius is the radius inside which the
mean density is 200 times the mean density of the Universe. The relation
between $\delta_c$ and $c$ is fixed by this definition. Unfortunately, the
definition of the virial radius is not unique. Often, the critical density of
the Universe instead of its mean density is referred to or overdensities
different from 200 are used. These differences have to be kept in mind when
comparing results from the NFW profile. 

Motivated by the Tully-Fisher and Faber-Jackson relations we assume a relation
between the virial radius and luminosity according to
$r_\mathrm{vir}/r_{\mathrm{vir},*}=(L/L_\star)^\eta$. As for the SIS, we adopt
$L_\star=10^{10}h^{-2}L_\odot$. 

\begin{figure}[ht!]
  \begin{center}
  \leavevmode
  \begin{minipage}[c]{0.49\textwidth}
    \includegraphics[width=\hsize,angle=-90]{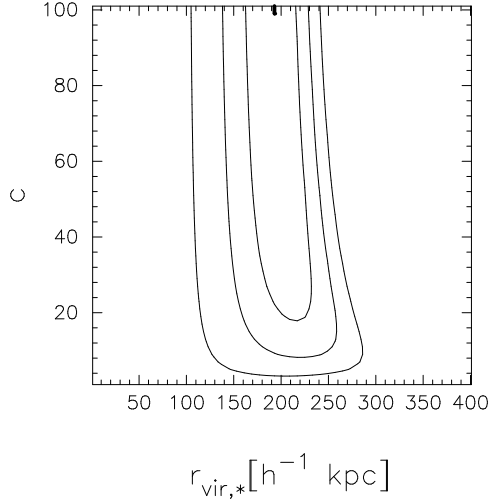}
  \end{minipage}
  \end{center}
  \caption{Constraints on dark matter halos modelled by NFW profiles,
    fitting virial radius $r_{\mathrm{vir},\star}$ and concentration $c$.}
  \label{nfw_c}
\end{figure}

First, we try to measure the virial radius $r_{\mathrm{vir},*}$ and the
concentration $c$ at fixed $\eta=0.3$, see Fig.\ \ref{nfw_c}. The virial
radius can be constrained well while on the concentration we can only derive
lower limits. This implies an upper limit on the scale radius,
$r_s<10h^{-1}~\mathrm{kpc}$. This is at all considered lens redshifts smaller
than the imposed minimum angular separation between lenses and sources of
8". Therefore, we cannot expect to be sensitive to $r_s$ or $c$. In the
following, we fix $c=20$ which is at the lower end of the values allowed by
our measurement. Note that when defining the virial radius as radius inside
which the mean density is 200 times the critical density of the Universe
(instead of its mean density as done here) this would refer to $c=12.5$.
Correspondingly, the virial radii and virial masses which we are going to
derive would be smaller in that case by about 40\% and 20\%, respectively.

\begin{figure}[ht!]
  \begin{center}
  \leavevmode
  \begin{minipage}[l]{0.49\textwidth}
    \includegraphics[width=\hsize,angle=-90]{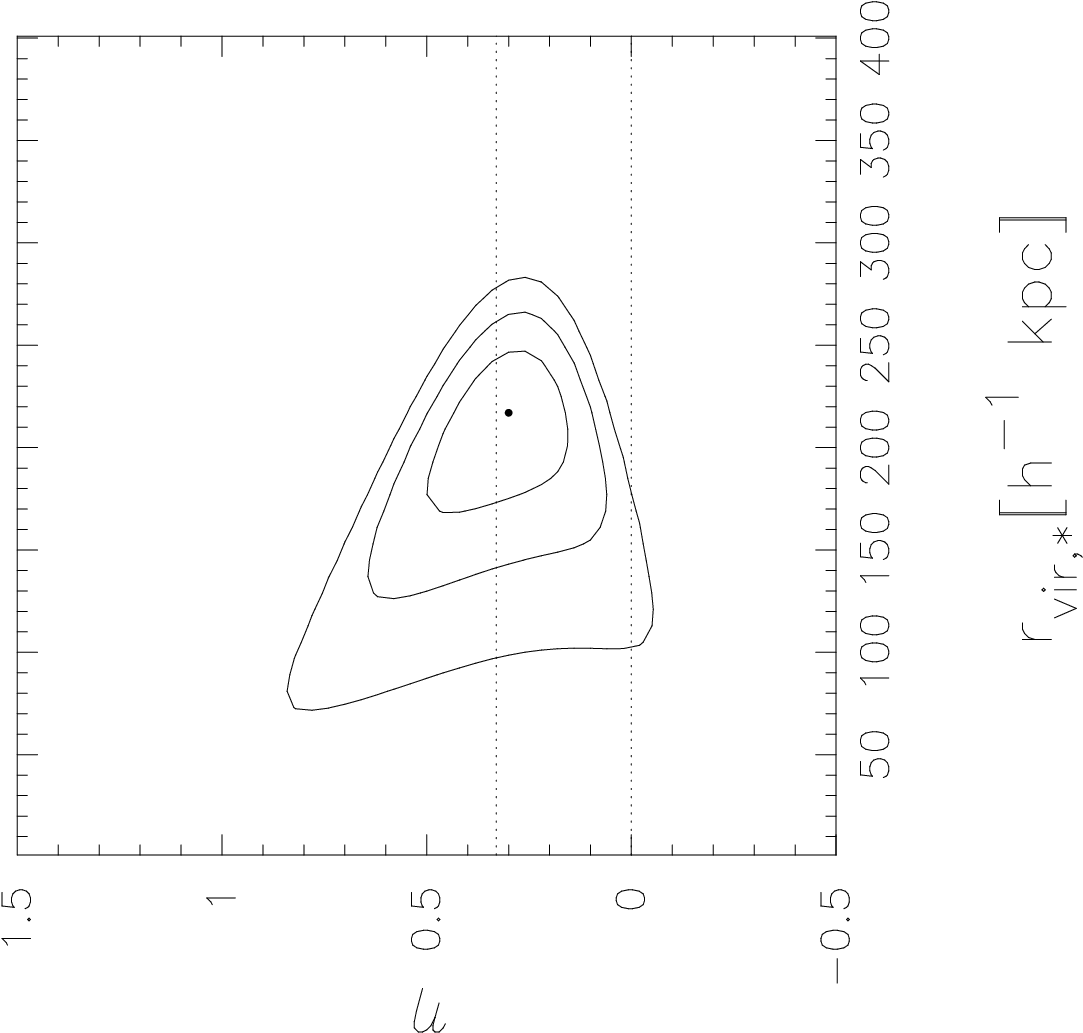}
  \end{minipage}
  \begin{minipage}[r]{0.49\textwidth}
    \includegraphics[width=\hsize,angle=-90]{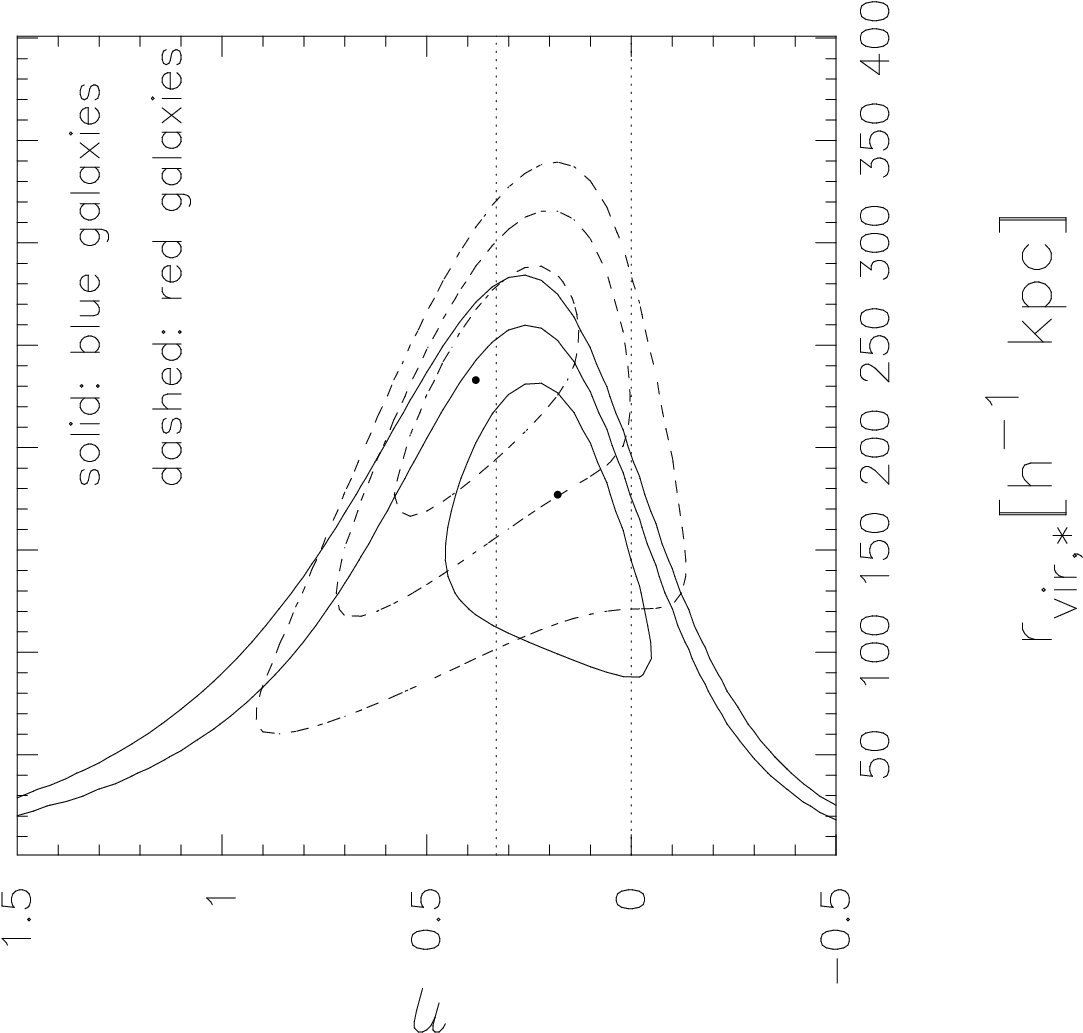}
  \end{minipage}
  \end{center}
  \caption{Constraints on dark matter halos modelled by NFW profiles, fitting
    virial radius $r_{\mathrm{vir},\star}$ and its dependence on luminosity
    ($\eta$). The left panel uses the whole lens sample. In the right panel,
    the lens sample is split into blue and red lenses.}
  \label{nfw}
\end{figure}

Figure \ref{nfw} shows 1-, 2- and 3-$\sigma$ contours for the virial radius
$r_{\mathrm{vir},*}$ and $\eta$ for the whole lens sample and for the blue and
red subsamples. Averaged over all lenses, the best-fit parameters with
1-$\sigma$ errors are $r_{\mathrm{vir},*}=217^{+24}_{-32}h^{-1}~\mathrm{kpc}$
and $\eta=0.30^{+0.16}_{-0.12}$. For the blue sample we find
$r_{\mathrm{vir},*}=177^{+40}_{-56}h^{-1}~\mathrm{kpc}$ and
$\eta=0.18^{+0.16}_{-0.16}$, for the red sample
$r_{\mathrm{vir},*}=233^{+48}_{-48}h^{-1}~\mathrm{kpc}$ and
$\eta=0.38^{+0.16}_{-0.20}$. Between the blue and red subsamples we measure a 1-$\sigma$ difference in the virial radius as well as in $\eta$.

\begin{table}\def~{\hphantom{0}}
  \begin{center}
  \caption{Constraints on dark matter halos of galaxies modelled by NFW
      profiles. The virial radius $r_{\mathrm{vir},*}$ and $\eta$ are fitted
      quantities (see Fig.\ \ref{nfw}), the virial mass
      $M_{\mathrm{vir},*}$, the virial mass-to-light ratio
      $M_{\mathrm{vir},*}/L$ and the rotation velocity at the virial radius,
      $v_{\mathrm{vir},*}$, are calculated from
      $r_{\mathrm{vir},*}$. $\beta=3\eta-1$ gives the scaling of
      $M_{\mathrm{vir},*}/L$ with luminosity, $M_{\mathrm{vir},*}/L\propto
      L^\beta$. The maximum rotation velocity $v_\mathrm{max}$ and the radius
      of the maximum rotation velocity $r(v_\mathrm{max})$ are calculated for
      a concentration $c=20$.}
  \begin{tabular}{lccccrrcc}\hline
     & $r_{\mathrm{vir},*}$ & $\eta$ & $M_{\mathrm{vir},*}$ &
     $M_{\mathrm{vir},*}/L$ & $\beta$ & $v_{\mathrm{vir},*}$ &
     $v_\mathrm{max}$ & $r(v_\mathrm{max})$\\
     & [$h^{-1}~\mathrm{kpc}$] & & [$10^{11}h^{-1}M_\odot$] & [$h(M/L)_\odot$]
     & & [$\mathrm{km/s}$] & [$\mathrm{km/s}$] & [$h^{-1}~\mathrm{kpc}$]\\\hline
 all & $217^{+24}_{-32}$ & $0.30^{+0.16}_{-0.12}$ & $7.1^{+2.6}_{-2.7}$ &
 $71^{+26}_{-27}$ & $-0.10^{+0.48}_{-0.36}$ & $119^{+13}_{-18}$ &
 $169^{+19}_{-25}$ & $23.4^{+2.6}_{-3.4}$\\
 blue & $177^{+40}_{-56}$ & $0.18^{+0.16}_{-0.16}$ & $3.9^{+3.3}_{-2.6}$ &
 $39^{+33}_{-26}$ & $-0.46^{+0.48}_{-0.48}$ & $97^{+22}_{-31}$ &
 $138^{+33}_{-44}$ & $19.1^{+4.3}_{-6.0}$\\
 red & $233^{+48}_{-48}$ & $0.38^{+0.16}_{-0.20}$ & $8.8^{+6.7}_{-4.4}$ &
 $88^{+67}_{-44}$ & $0.14^{+0.62}_{-0.60}$ & $128^{+26}_{-26}$ &
 $181^{+38}_{-36}$ & $25.2^{+5.1}_{-5.2}$\\\hline
  \end{tabular}
 \end{center}
 \label{table}
\end{table}
Table \ref{table} gives an overview of the measured parameters
($r_{\mathrm{vir},*}$, $\eta$) and calculated quantities like the virial
mass $M_{\mathrm{vir},*}$, virial mass-to-light ratio $M_{\mathrm{vir},*}/L$
and the scaling between $M_{\mathrm{vir},*}/L$ and luminosity.

Again, we compare our results to those from other data
sets. \cite{hoekstra2004} find from the RCS
$M_{\mathrm{vir},*}=8.4\pm0.7\times10^{11}h^{-1}M_\odot$ at
$L_B=10^{10}h^{-2}L_\odot$. At the corresponding $L_\star=1.25\times
10^{10}h^{-2}L_\odot$ measured in the restframe $r$-band we find
$M_{\mathrm{vir},*}=8.0^{+3.9}_{-3.0}\times10^{11}h^{-1}M_\odot$.
\cite{guzik2002} obtain
$M_{\mathrm{vir},*}=8.96\pm1.59\times10^{11}h^{-1}M_\odot$ at
$L_\star=1.51\times 10^{10}h^{-2}L\odot$ from the SDSS where $L_\star$ is
measured in the SDSS restframe $r$-band. However,
\cite{guzik2002} define the virial radius as radius inside which the mean
density is 200 times the critical density of the Universe. Our corresponding
result using their value of $L_\star$ and their definition of the virial
radius is $M_{\mathrm{vir},*}=7.8^{+3.5}_{-2.7}\times10^{11}h^{-1}M_\odot$. 
 
\subsection{Individual fields}
Finally, we investigate the three survey fields used here individually and
address the question whether they give consistent results. 
\begin{figure}[ht!]
  \begin{center}
  \leavevmode
  \begin{minipage}[l]{0.32\textwidth}
    \includegraphics[width=\hsize,angle=-90]{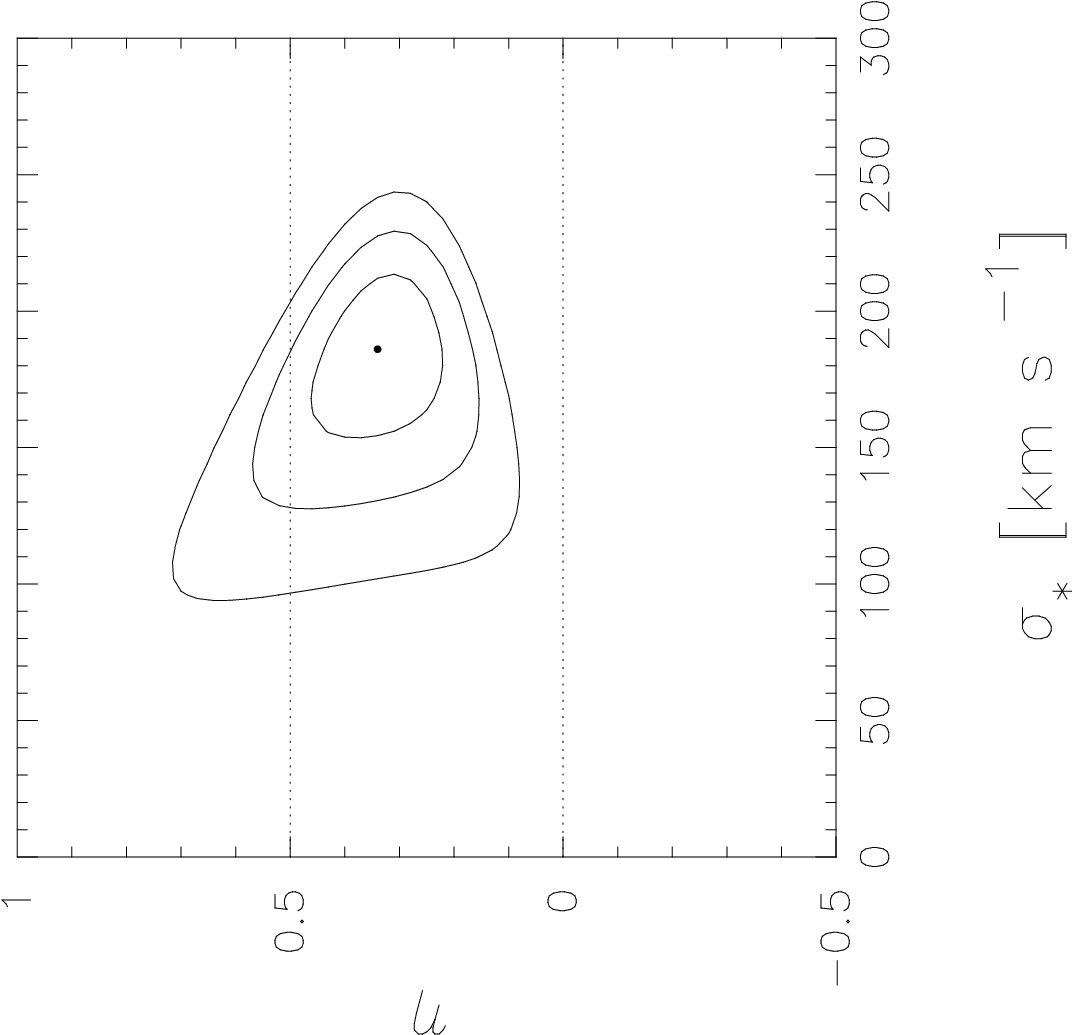}
  \end{minipage}
  \begin{minipage}[c]{0.32\textwidth}
    \includegraphics[width=\hsize,angle=-90]{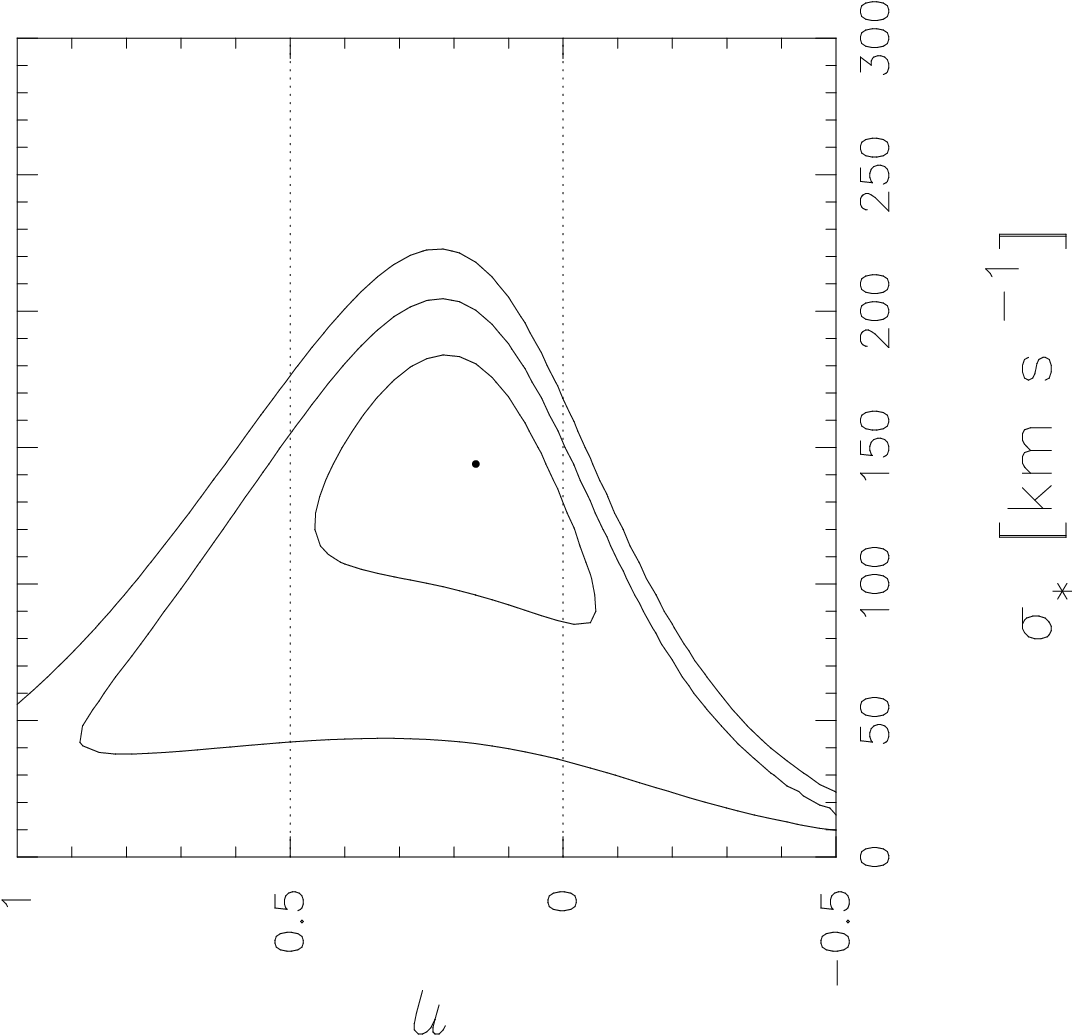}
  \end{minipage}
  \begin{minipage}[r]{0.32\textwidth}
    \includegraphics[width=\hsize,angle=-90]{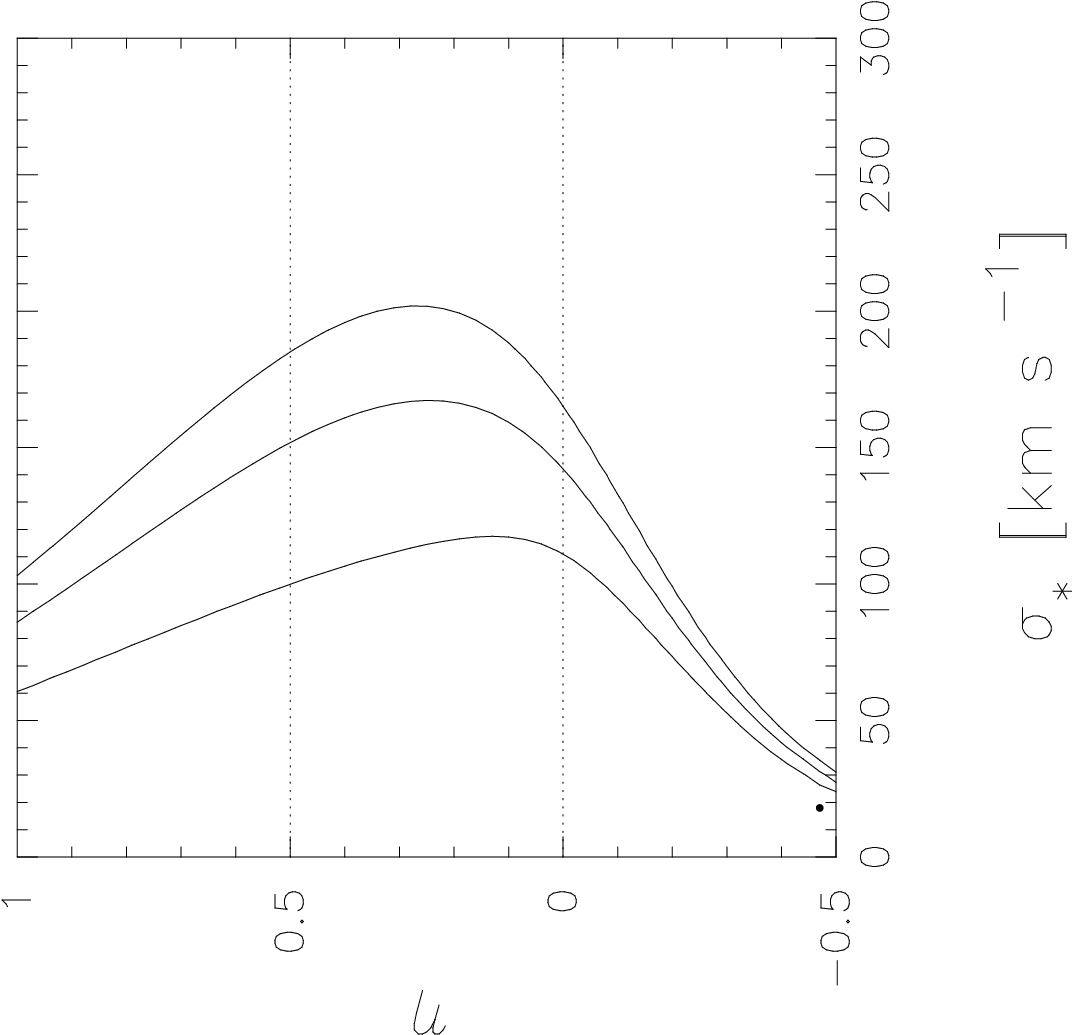}
  \end{minipage}
  \end{center}
  \caption{Constraints on dark matter halos modelled as SIS from the three
    individual survey fields. The left panel refers to the A901 field, the
    middle panel to the S11 field and the right panel to the CDFS field.}
  \label{fields}
\end{figure}
Figure \ref{fields} shows likelihood contours from fitting the SIS model as in
Sect.\ \ref{sect:sis}. Clearly, the derived constraints on the velocity
dispersion $\sigma_\star$ are not consistent for the individual fields. While
the A901 field gives very tight constraints, from the CDFS we can only derive
an upper limit on $\sigma_\star$. The deviation from the measurement that uses
all three fields together ($\sigma_\star=156~\mathrm{km/s}$) is 1-$\sigma$
towards higher values for the A901 field and 2-$\sigma$ towards lower values
for the CDFS field. Only the S11 field is consistent with the overall
measurement.

These three survey fields were already selected to be very different. The
S11 field is the only random field. It contains the cluster Abell 1364 at
$z=0.11$ by chance. The A901 field was chosen because it contains a
supercluster with the three components Abell 901a, 901b and 902 at
$z=0.16$. The CDFS field contains the Chandra Deep Field South and was chosen
because of its emptiness. Due to these selection criteria one might suspect
that the measured galaxy-galaxy lensing signal is mostly due the foreground
clusters. However, our lens sample only uses galaxies at $z>0.2$ and should
thus not include cluster galaxies. Therefore, the foreground clusters in the
A901 and S11 field should not have any influence on our measurement. By
including the additional shear from the foreground clusters we indeed
confirm this assumption. Further, we find that the shear from an additional
cluster in the A901 field at $z=0.47$ does not induce a significant shift in
the velocity dispersion. Including the shear from this higher-redshift cluster
does however increase the uncertainties by about 20\%.

Another suspected reason for the differences between the three fields is the
imaging quality. The sum image of the A901 field has the best quality with a
PSF of 0.74". The PSF of the other two sum images is 0.88". If image quality
had a dominant effect, then the results from the S11 field and the CDFS field
should be consistent. For the CDFS field we have several independent sum
images available from different observing runs with very different seeing
conditions and exposure times which yield consistent lensing
signals. Therefore we rule out image quality as possible explanation for the
discrepant measurements.

The most probable explanation we find for the deviating results comes from the
number counts in the different fields. The number of lenses in the fields is
4636 (A901), 4268 (S11) and 3573 (CDFS). Therefore, one expects the tightest
constraints from the A901 field. The difference in the derived velocity
dispersion could be due to differences in the composition of the lens
samples. Indeed, the fraction of red lenses is 23.5\% in the A901 field,
20.5\% in the S11 field and only 17.2\% in the CDFS field. Given our findings
in Sect.\ \ref{sect:sis} we expect a higher velocity dispersion and tighter
constraints with increasing fraction of red galaxies.


\end{document}